\documentclass[12pt]{iopart}
\usepackage{iopams}  
\usepackage{graphicx}

\eqnobysec

\begin{document}

\newtheorem{proposition}{Proposition}[section]

\newcommand\mythesection{\arabic{section}}

\newcommand\mynumparts[1]{\refstepcounter{equation}%
   \if ?#1?\else\label{#1}\fi%
   \setcounter{eqnval}{\value{equation}}%
   \setcounter{equation}{0}%
   \def\theequation{\mythesection.\arabic{eqnval}{\it\alph{equation}}}%
   }

\def\myendnumparts{
   \def\theequation{\mythesection.\arabic{equation}}%
   \setcounter{equation}{\value{eqnval}}%
   }

\renewcommand\kappa{\varkappa} 
\renewcommand\ge{\geqslant} 
\renewcommand\le{\leqslant} 
\renewcommand\vec[1]{\boldsymbol{#1}}

\newcommand\mynode[1]{\mathsf{#1}}
\newcommand\myvar[1]{\mathop{#1}} 
\newcommand\myin[1]{\vec{n} \in \Lambda^{#1}}
\newcommand\myphip{\vec{\phi}_{+}}
\newcommand\myphin{\vec{\phi}_{-}}
\newcommand\myPhip{\vec{\varphi}^{+}}
\newcommand\myPhin{\vec{\varphi}^{-}}

\newcommand{\myauxspace}{\mathbb{V}}
\newcommand{\mybasevec}{\vec{x}}

\newcommand{\mychi}[1]{\chi_{#1}}
\newcommand{\mychis}[2]{\mathop{\chi_{#1,#2}}}

\newcommand{\myconst}[2]{\mathop{#1_{#2}}}

\newcommand\myShift[2][T]{\mathbb{#1}_{#2}}
\newcommand\myShifted[3][T]{\mathbb{#1}_{#2}{#3}} 
\newcommand\myShiftedB[3][T]{\left(\mathbb{#1}_{#2}{#3}\right)} 
\newcommand\myShiftedInv[3][T]{\mathbb{#1}_{#2}^{-1}{#3}} 
\newcommand\myShiftedInvB[3][T]{\left(\mathbb{#1}_{#2}^{-1}{#3}\right)} 

\newcommand{\mytauU}{\rho^{(3)}} 
\newcommand{\mytauV}{\sigma^{(3)}} 
\newcommand{\mytauQ}{\sigma^{(2)}} 
\newcommand{\mytauR}{\rho^{(2)}} 

\newcommand\mymatrix[1]{\mathsf{#1}} 
\newcommand\myunitket{|\,1\,\rangle}  
\newcommand\mybra[1]{\langle\,#1\,|} 

\newcommand{\myhalf}{\mbox{\small $\frac{1}{2}$}} 
\newcommand{\myslambda}{\vec{\lambda}_{\alpha}} 
\newcommand{\mysmu}{\mu_{\alpha}} 
\newcommand{\mysphase}[1]{\theta_{#1}(\vec{n})} 
\newcommand{\mysconst}[1]{\delta_{#1}} 


\title{Solitons of a vector model on the honeycomb lattice.}
\author{V.E. Vekslerchik}
\address{
  Usikov Institute for Radiophysics and Electronics \\
  12, Proskura st., Kharkov, 61085, Ukraine 
}
\ead{vekslerchik@yahoo.com}
\submitto{\JPA}

\begin{abstract}
We study a simple nonlinear vector model defined on the honeycomb lattice. 
We propose a bilinearization scheme for the field equations and demonstrate 
that the resulting system is closely related to the well-studied integrable 
models, such as the Hirota bilinear difference equation and the 
Ablowitz-Ladik system. This result is used to derive the $N$-soliton 
solutions. 
\end{abstract}

\section{Introduction.}

We study a simple nonlinear model defined on the honeycomb lattice (HL). 
The main aim of this work is to apply the direct methods of the soliton 
theory to the case of the `non-square', i.e. different from $\mathbb{Z}^{2}$, 
two-dimensional lattices.

There has been considerable interest in the integrable nonlinear 
models on such lattices, and even arbitrary graphs (see, for example, 
\cite{A98,A01,BS02,BHS02,BH03,AS04,DNS07,BS10}).  
The results of these studies provide answers to many questions arising in the 
theory of integrable systems. However, if we consider the problem of finding 
solutions, there is still, in our opinion, much to be done in this field. 
The case is that many of the standard tools have not been adapted so far 
to the non-square lattices. For example, for many models on graphs 
(in particular, as is shown in \cite{BS02}, for all models that possess the 
property of the three-dimensional consistency \cite{ABS03}) 
one can construct a special form of the Lax (or zero-curvature) representation, 
called the ``trivial monodromy representation'' in \cite{A01}, which has been 
successively used as an integrability test. 
Nevertheless, the graph analogue of the inverse scattering transform (IST), 
that is based on this representation, has not been elaborated yet. 
In this situation, the main tool to derive explicit solutions are the 
so-called direct methods, for which the lack of natural ways to separate 
variables (as in the case of HL) seems to be less important than for the 
IST-like approaches. 
Of course, to make these methods suitable for the HL, 
one has to modify the standard procedure.
However, as a reader will see, 
this can be done by rather elementary means.

In this work we present the explicit $N$-soliton solutions for the vector 
model which is described in section \ref{sec:model}. 
In section \ref{sec:bilin}, we bilinearize the field equations and convert 
them into a simple system of three-point equations.  
In section \ref{sec:hal}, we discuss this system, and show that it is closely 
related to the well-studied integrable models, such as the Hirota bilinear 
difference and the Ablowitz-Ladik equations. 
Then, using the already known results as well as the ones derived in section 
\ref{sec:hal}, we present, in section \ref{sec:sls}, the $N$-soliton solutions 
for the field equations of our model.

\section{The model and the main equations. \label{sec:model}}

\begin{figure}%
\begin{center}
\includegraphics{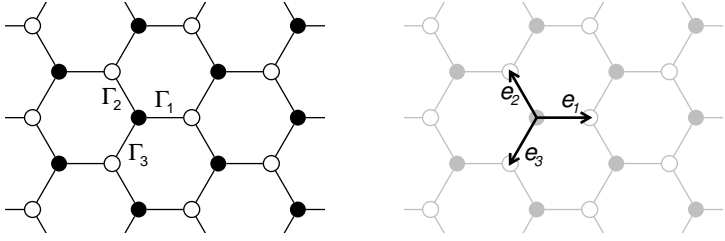}%
\end{center}
\caption{
  Bipartition of the HL, interaction constants and base vectors.
  The vertices that belong to $\Lambda^{+}$ are shown 
  by black circles and the vertices that belong to $\Lambda^{-}$ 
  are shown by white ones.
}%
\label{fig-1}%
\end{figure}

The model which we study in this paper describes the 
three-dimensional vectors (fields) 
$\vec\phi=\vec\phi(\mynode{v}) \in \mathbb{R}^{3}$ 
defined at the vertices $\mynode{v}$ of the HL with the logarithmic 
interaction between the nearest neighbours,
\begin{equation}
	\mathcal{S} 
  = 
  \sum\limits_{ \mynode{v'}\sim\mynode{v''} } 
  \Gamma_{(\mynode{v'},\mynode{v''})} \; 
  \ln\left[ 
    1 + \left(\vec{\phi}(\mynode{v'}), \vec{\phi}(\mynode{v''}) \right) 
  \right] 
\label{def:SNN}
\end{equation}
where the notation $\mynode{v'}\sim\mynode{v''}$ means that the vertices 
$\mynode{v'}$ and $\mynode{v''}$ are connected by an edge of the HL, 
$\left( \vec\phi', \vec\phi'' \right)$ is the standard scalar product 
in $\mathbb{R}^{3}$ and 
$\Gamma_{(\mynode{v'},\mynode{v''})}$ 
are constants 
which take the values $\Gamma_{1}$, $\Gamma_{2}$ or $\Gamma_{3}$ depending on the 
\emph{direction} of the edge $(\mynode{v'},\mynode{v''})$ connecting nodes 
$\mynode{v'}$ and $\mynode{v''}$ 
(see figure \ref{fig-1})
and satisfy the following restriction: 
\begin{equation}
	\sum_{\mynode{v}'} 
  \Gamma_{(\mynode{v},\mynode{v'})} 
  = 0 
  \quad
  \mbox{for all} \; \mynode{v} 
\label{restr:gvv}
\end{equation}
with the summation over all nodes adjacent to $\mynode{v}$.

The logarithmic interaction in \eref{def:SNN} between the vectors $\vec\phi$ 
is not new to the theory of integrable systems (see, for example, 
\cite{I82,H82,P87}) and can be viewed as the classical$+$integrable analogue 
of the famous Heisenberg interaction of the quantum mechanics. In this sense, 
the model considered here is closely related to the one-dimensional 
Ishimori spin chain \cite{I82}. 
However, there is an essential difference: we \emph{do not impose restrictions} 
like $\vec\phi^{2}=1$ which are crucial for models describing 
the spin-like systems.  

On the other hand, this model can be considered as a vector generalization 
of one of the `universal' integrable models of the paper \cite{A01} which was 
studied in \cite{BS02,V16}. 

Considering the condition \eref{restr:gvv}, it should be noted that 
restrictions of this type often appear in the studies of integrable models. 
If we, for example, look at the Hirota bilinear difference equation (HBDE), 
the restriction similar to \eref{restr:gamma} is present in the most of the 
works devoted to this system (including the original paper \cite{H81}). 
However, as it has been demonstrated in, for example, \cite{RGS92}, it is not 
needed for the integrability (it is a widespread opinion that it is required 
for the existence of the Hirota-form soliton solutions).

Hereafter, we use the vector notation. We introduce coplanar vectors 
$\vec{e}_{1}$, $\vec{e}_{2}$ and $\vec{e}_{3}$, 
that generate the HL and are related by 
\begin{equation}
	 \sum_{i=1}^{3} \vec{e}_{i} = \vec{0}, 
\label{e-restriction}
\end{equation}
the set $\Lambda$ of the lattice vectors $\vec{n}$ 
(positions of the vertices of the HL), 
\begin{equation}
  \Lambda 
  = 
  \left\{\left. 
    \vec{n} = \sum_{i=1}^{3} n_{i} \vec{e}_{i}, 
    \quad
    n_{i} \in \mathbb{Z}
    \quad \right| \quad
    \sum_{i=1}^{3} n_{i} \ne 2 \mathop{\mbox{mod}} 3
  \right\},
\end{equation}
which can be decomposed as 
\begin{equation}
 \Lambda = \Lambda^{+} \cup \Lambda^{-} 
\end{equation} 
with 
\begin{equation}
  \begin{array}{l}
  \Lambda^{+} 
  = 
  \left\{ \left. 
    \vec{n} = \sum\limits_{i=1}^{3} n_{i} \vec{e}_{i}, 
    \quad
    n_{i} \in \mathbb{Z}
    \quad \right| \quad
    \sum\limits_{i=1}^{3} n_{i} = 0 \mathop{\mbox{mod}} 3 \; 
  \right\}, 
  \\[4mm] 
  \Lambda^{-} 
  = 
  \left\{ \left. 
    \vec{n} = \sum\limits_{i=1}^{3} n_{i} \vec{e}_{i}, 
    \quad
    n_{i} \in \mathbb{Z}
    \quad \right| \quad
    \sum\limits_{i=1}^{3} n_{i} = 1 \mathop{\mbox{mod}} 3 \; 
  \right\} 
  \end{array}
\end{equation}
(this is a manifestation of the fact that the HL is a bipartite graph) 
and write $\vec{\phi}(\vec{n})$ instead of $\vec{\phi}(\mynode{v})$.

In the $\vec{n}$-terms the action \eref{def:SNN} can be presented as 
\mynumparts{def:Spn}
\begin{eqnarray}
  \mathcal{S} 
  & = & 
  \sum\limits_{\myin{+}} 
  \sum\limits_{i=1}^{3} 
    \Gamma_{i} 
    \ln\left[ 
      1 + 
      \left(\vec{\phi}(\vec{n}), \vec{\phi}(\vec{n}+\vec{e}_{i}) \right) 
    \right] 
  \\
  & = & 
  \sum\limits_{\myin{-}} 
  \sum\limits_{i=1}^{3} 
    \Gamma_{i} 
    \ln\left[ 
      1 + 
      \left(\vec{\phi}(\vec{n}), \vec{\phi}(\vec{n}-\vec{e}_{i}) \right) 
    \right] 
\label{def:SPN}
\end{eqnarray}
\myendnumparts
where we use, instead of $\Gamma_{(\mynode{v'},\mynode{v''})}$, 
constants $\Gamma_{i}$ ($i=1,2,3$), 
$
  \Gamma_{(\mynode{v'},\mynode{v''})} 
  = 
  \Gamma_{i} 
$
if the edge $(\mynode{v'},\mynode{v''})$ is parallel to the vector 
$\vec{e}_{i}$ (see figure \ref{fig-1}), 
subjected to the restriction \eref{restr:gvv}, 
\begin{equation}
    \sum\limits_{i=1}^{3} \Gamma_{i} = 0.
\label{restr:gamma}
\end{equation}
The `variational' equations 
\begin{equation}
  \left. \partial \mathcal{S} \right/ \partial \vec{\phi}(\vec{n}) = 0, 
  \qquad
  \vec{n} \in \Lambda 
\label{eq-variation}
\end{equation}
can be written as 
\mynumparts{syst:main}
\begin{eqnarray}
  \sum\limits_{i=1}^{3} 
  \frac{ \Gamma_{i} }
       { 1 + 
         \left(\vec{\phi}(\vec{n}), \vec{\phi}(\vec{n}+\vec{e}_{i}) \right) 
       }\; 
  \vec{\phi}(\vec{n} + \vec{e}_{i}) 
  = \vec{0}
  &\qquad& 
  ( \myin{+} )
  \\[2mm]
  \sum\limits_{i=1}^{3} 
  \frac{ \Gamma_{i} }
       { 1 + 
         \left(\vec{\phi}(\vec{n}), \vec{\phi}(\vec{n}-\vec{e}_{i}) \right) 
       }\; 
  \vec{\phi}(\vec{n} - \vec{e}_{i}) 
  = \vec{0}
  && 
  ( \myin{-} )
\end{eqnarray}
\myendnumparts
Namely these equations are the main object of our study.

\section{Solving the field equations. \label{sec:bilin}}

In this section we reduce the field equations \eref{syst:main} to an already 
known bilinear system. 
The procedure, which is, for the most part, rather standard has a few 
non-trivial moments that stem from the structure of the HL. 

\subsection{Resolving the restriction \eref{e-restriction}.}

To resolve the restriction \eref{e-restriction} 
we, first, `replace' the vectors $\vec{e}_{i}$ which obey 
\eref{e-restriction} with new arbitrary vectors 
$\vec{\alpha}_{i}$  ($i=1,2,3$) from some auxiliary space 
$\myauxspace$. 
This means that we consider instead of functions of 
$\vec{n}=\sum_{i=1}^{3} n_{i} \vec{e}_{i}$, 
functions of 
$\sum_{i=1}^{3} n_{i} \left( \vec{\alpha}_{i} - \vec{\delta} \right)$  
with 
$
  \vec{\delta} 
  = 
  \frac{1}{3} \sum_{i=1}^{3} \vec{\alpha}_{i}. 
$
Thus, we introduce the map $\mybasevec: \Lambda \to \myauxspace$, 
\begin{equation}
  \vec{n} 
  = 
  \sum_{i=1}^{3} n_{i} \vec{e}_{i} 
  \quad \to \quad 
  \mybasevec(\vec{n}) 
  = 
  \sum_{i=1}^{3} n_{i} \left( \vec{\alpha}_{i} - \vec{\delta} \right), 
  \quad 
  \vec{\delta} 
  = 
  \frac{1}{3} \sum_{i=1}^{3} \vec{\alpha}_{i} 
\label{def:nu}
\end{equation}
whose image belongs to the two-dimensional 
plane from our auxiliary space $\myauxspace$.  
The advantage of the $\mybasevec$-representation is that it
automatically (for arbitrary $\vec{\alpha}_{i}$) takes into account the 
restriction \eref{e-restriction}, 
$ 
  \mybasevec\left( \vec{n} + \sum_{i=1}^{3} \vec{e}_{i} \right) 
  = 
  \mybasevec(\vec{n})
$. 

Secondly, in order to simplify the following equations 
and eliminate the explicit appearance of $\vec{\delta}$, 
we introduce, instead of the vectors $\vec{\phi}$, new vectors, 
$\myphip$ and $\myphin$,  
\begin{equation}
  \vec{\phi}(\vec{n}) 
  = 
  \left\{ 
  \begin{array}{rcl} 
    \myphip(\mybasevec(\vec{n}) + \vec{\delta}) && ( \myin{+} ) 
    \\
    \myphin(\mybasevec(\vec{n}) - \vec{\delta}) && ( \myin{-} ). 
  \end{array}
  \right.
\label{def:qr}
\end{equation}
In new terms, we can rewrite the equations we want to solve as 
\mynumparts{syst:lattice-qr}
\begin{eqnarray}
\fl\; 
  \sum\limits_{i=1}^{3} 
  \frac{ \Gamma_{i} }
       { 1 + 
         \left(\myphip(\mybasevec_{+}), 
               \myphin(\mybasevec_{+} - \vec{\alpha}_{i-1} - \vec{\alpha}_{i+1}) 
         \right) 
       }\; 
  \myphin(\mybasevec_{+} - \vec{\alpha}_{i-1} - \vec{\alpha}_{i+1}) 
  = \vec{0}
  &\;& 
  ( \myin{+} )
  \\[2mm]
\fl\; 
  \sum\limits_{i=1}^{3} 
  \frac{ \Gamma_{i} }
       { 1 + 
         \left( \myphip(\mybasevec_{-} + \vec{\alpha}_{i-1} + \vec{\alpha}_{i+1}), 
                \myphin(\mybasevec_{-}) \right) 
       }\; 
  \myphip(\mybasevec_{-} + \vec{\alpha}_{i-1} + \vec{\alpha}_{i+1}) 
  = \vec{0}
  && 
  ( \myin{-} )
\end{eqnarray}
\myendnumparts
where $\mybasevec_{\pm} = \mybasevec(\vec{n}) \pm \vec\delta$. 
In these  equations, as well as in the rest of the paper, 
we use the following convention: all arithmetic operations with 
$\vec\alpha$- and $\Gamma$-indices are understood modulo $3$,
\begin{equation}
  \vec\alpha_{i \pm 3} = \vec\alpha_{i}, 
  \qquad 
  \Gamma_{i \pm 3} = \Gamma_{i}
  \qquad
  (i=1,2,3).
\end{equation}

Looking at \eref{def:qr} one can see that the `natural' domains of definition 
of the functions $\myphip$ and $\myphin$ are points of the lattices 
$\mybasevec(\Lambda^{\pm}) \pm \vec\delta$ (which belong to parallel, but 
different, 2-planes of $\myauxspace$). 
However, we consider both $\myphip$ and $\myphin$ as defined on the whole 
$\myauxspace$ ($ \myphip,\myphin: \myauxspace \to \mathbb{R}^{3} $) and 
\emph{define} them as solutions of the system similar to 
\eref{syst:lattice-qr} but thought of as a system on $\myauxspace$: 
\begin{equation}
  \left\{
  \begin{array}{l}
  \sum\limits_{i=1}^{3} 
  \Gamma_{i} \, H_{i}^{+}(\mybasevec) \; 
  \myphin(\mybasevec - \vec{\alpha}_{i-1} - \vec{\alpha}_{i+1}) 
  =  
  \vec{0}
  \\
  \sum\limits_{i=1}^{3} 
  \Gamma_{i} \, H_{i}^{-}(\mybasevec) \; 
  \myphip(\mybasevec + \vec{\alpha}_{i-1} + \vec{\alpha}_{i+1}) 
  = 
  \vec{0}
  \end{array}
  \right.
  \qquad
  (\mybasevec \in \myauxspace) 
\label{syst:main-hqr} 
\end{equation}
where 
\mynumparts{def:Hpm}
\begin{eqnarray} 
  H_{i}^{+}(\mybasevec) 
  & = & 
  \left[ 
  1 + 
  \left(
    \myphip(\mybasevec), 
    \myphin(\mybasevec - \vec{\alpha}_{i-1} - \vec{\alpha}_{i+1}) 
  \right) 
  \right]^{-1}, 
\\[2mm]   
  H_{i}^{-}(\mybasevec) 
  & = & 
  \left[
  1 + 
  \left( 
    \myphip(\mybasevec + \vec{\alpha}_{i-1} + \vec{\alpha}_{i+1}), 
    \myphin(\mybasevec) 
  \right) 
  \right]^{-1}. 
\end{eqnarray}
\myendnumparts

\subsection{Ansatz.}

The key step of our construction is the following \textit{ansatz}: 
\begin{equation}
\begin{array}{lcl}
  \myphip(\mybasevec + \vec{\alpha}_{j} + \vec{\alpha}_{k}) 
  & \propto & 
  \myphip(\mybasevec + \vec{\alpha}_{j} ) 
  - 
  \myphip(\mybasevec + \vec{\alpha}_{k}), 
  \\ 
  \myphin(\mybasevec - \vec{\alpha}_{j} - \vec{\alpha}_{k}) 
  & \propto & 
  \myphin(\mybasevec - \vec{\alpha}_{j} ) 
  - 
  \myphin(\mybasevec - \vec{\alpha}_{k}). 
  \end{array}
\end{equation}
Of course, this ansatz is rather restrictive. However, it leads to a rather 
wide range of solutions, which include the $N$-soliton solutions (as well as 
the so-called finite-gap, Toeplitz and other solutions). 

In more detail, we require 
\mynumparts{def:ansatz} 
\begin{eqnarray} 
\fl\qquad
  H_{i}^{+}(\mybasevec) \; 
  \myphin(\mybasevec - \vec{\alpha}_{i-1} - \vec{\alpha}_{i+1}) 
  & = & 
  \frac{
    \myphin(\mybasevec - \vec{\alpha}_{i-1} ) 
    - 
    \myphin(\mybasevec - \vec{\alpha}_{i+1}) 
  }{ \mychi{i-1} - \mychi{i+1} }, 
\\ 
\fl\qquad
  H_{i}^{-}(\mybasevec) \; 
  \myphip(\mybasevec + \vec{\alpha}_{i-1} + \vec{\alpha}_{i+1}) 
  & = & 
  \frac{ 
    \myphip(\mybasevec + \vec{\alpha}_{i-1} ) 
    - 
    \myphip(\mybasevec + \vec{\alpha}_{i+1}) 
  }{ \mychi{i-1} - \mychi{i+1} }
\end{eqnarray}
\myendnumparts
with constants $\mychi{i}$ ($i=1,2,3$) which will be specified below. Again, 
we presume that $\mychi{i \pm 3} = \mychi{i}$.

With \eref{def:ansatz}, equations \eref{syst:main-hqr} become 
\mynumparts{syst:aqr} 
\begin{eqnarray}
  \sum\limits_{i=1}^{3} 
  \hat\Gamma_{i} \, 
  \myphin(\mybasevec - \vec{\alpha}_{i} ) 
  = 
  \vec{0}, 
\\
  \sum\limits_{i=1}^{3} 
  \hat\Gamma_{i} \, 
  \myphip(\mybasevec + \vec{\alpha}_{i}) 
  = 
  \vec{0}
\end{eqnarray}
\myendnumparts 
with \emph{constant} $\hat\Gamma_{i}$, 
\begin{equation}
  \hat\Gamma_{i} 
  = 
  \frac{ \Gamma_{i+1} }{ \mychi{i} - \mychi{i-1} } 
  + 
  \frac{ \Gamma_{i-1} }{ \mychi{i} - \mychi{i+1} }. 
\end{equation}
It is easily seen that we can satisfy all equations \eref{syst:aqr} 
without imposing additional conditions upon the vectors $\myphip$ and $\myphin$ 
by making all $\hat\Gamma_{i}$ equal to zero. 
Solution of this elementary problem leads to the following restriction upon 
the constants $\mychi{i}$ that can be 
used in the \textit{ansatz} \eref{def:ansatz}: 
\begin{equation}
  \sum_{i=1}^{3} \Gamma_{i} \, \mychi{i} = 0
\label{restr:chi}
\end{equation} 
(it is straightforward to verify that \eref{restr:chi} indeed leads to  
$\hat\Gamma_{i}=0$). 
Thus, we have reduced our problem to equations \eref{def:ansatz}, 
\eref{def:Hpm} together with \eref{restr:chi}.

\subsection{Bilinerization.}

Noting that $H_{i}^{\pm}$, 
considered as functions on $\myauxspace$, 
are related by simple shifts and using only one of 
them, 
\begin{equation}
  H_{i}(\mybasevec) := H_{i}^{-}(\mybasevec), 
\quad
  H_{i}^{+}(\mybasevec) 
  = 
  H_{i}(\mybasevec - \vec{\alpha}_{i-1} - \vec{\alpha}_{i+1}) 
\end{equation}
one arrives at the system 
\mynumparts{syst:qr}
\begin{eqnarray}
\fl\quad 
  \left( \mychi{i-1} - \mychi{i+1} \right)
  H_{i}(\mybasevec) \; 
  \myphip(\mybasevec + \vec{\alpha}_{i-1} + \vec{\alpha}_{i+1}) 
  & = & 
  \myphip(\mybasevec + \vec{\alpha}_{i-1} ) 
  - 
  \myphip(\mybasevec + \vec{\alpha}_{i+1}), 
\\
\fl\quad 
  \left( \mychi{i-1} - \mychi{i+1} \right)
  H_{i}(\mybasevec) \; 
  \myphin(\mybasevec) 
  & = & 
  \myphin(\mybasevec + \vec{\alpha}_{i+1} ) 
  - 
  \myphin(\mybasevec + \vec{\alpha}_{i-1}) 
\end{eqnarray}
and 
\begin{equation}
\fl\qquad 
  H_{i}(\mybasevec) 
  = 
  \left[ 
    1 + 
    \left( 
    \myphip(\mybasevec + \vec{\alpha}_{i-1} + \vec{\alpha}_{i+1}), 
    \myphin(\mybasevec) 
    \right) 
  \right]^{-1}. 
\end{equation}
\myendnumparts
This system can be easily bilinearized by introducing the tau-functions 
$\tau(\mybasevec)$ by 
\begin{equation}
  H_{i}(\mybasevec) 
  = 
  \mychis{i-1}{i+1} \; 
  \frac{ 
    \tau(\mybasevec) \tau(\mybasevec + \vec{\alpha}_{i-1} + \vec{\alpha}_{i+1}) 
  }{ 
    \tau(\mybasevec + \vec{\alpha}_{i-1}) \tau(\mybasevec + \vec{\alpha}_{i+1}) 
  } 
\end{equation}
with \emph{arbitrary} symmetric constants $\mychis{j}{k}$ 
($\mychis{j}{k} = \mychis{k}{j}$ 
and
$\mychis{j}{k} = \mychis{j \pm 3}{k} = \mychis{j}{k \pm 3}$) 
together with the vector tau-functions 
$\vec\sigma(\mybasevec)$ and $\vec\rho(\mybasevec)$ defined by 
\begin{equation}
  \myphip(\mybasevec) = \vec\sigma(\mybasevec) / \tau(\mybasevec), 
  \qquad 
  \myphin(\mybasevec) = \vec\rho(\mybasevec) / \tau(\mybasevec). 
\end{equation}

To summarize, the main result of this section can be formulated as 
%
\begin{proposition}
A wide range of solutions for the field equations \eref{syst:main} 
can be obtained by 
\begin{equation}
  \vec{\phi}(\vec{n}) 
  = 
  \left\{ 
  \begin{array}{rcl} 
    \vec\sigma(\mybasevec(\vec{n}) + \vec{\delta}) / 
    \tau(\mybasevec + \vec{\delta}) 
    && ( \myin{+} ) 
    \\
    \vec\rho(\mybasevec(\vec{n}) - \vec{\delta}) / 
    \tau(\mybasevec - \vec{\delta}) 
    && ( \myin{-} ) 
  \end{array}
  \right.
\end{equation}
where $\mybasevec(\vec{n})$ and $\vec\delta$ are defined in \eref{def:nu}, 
$\vec\sigma(\mybasevec)$, $\vec\rho(\mybasevec)$ and $\tau(\mybasevec)$ 
are solutions for the system 
\mynumparts{syst:bilin}
\begin{eqnarray}
&&
  \left( \mychi{j} - \mychi{k}\right) 
  \mychis{j}{k} \; 
  \tau(\mybasevec) 
  \vec\sigma(\mybasevec + \vec{\alpha}_{i-1} + \vec{\alpha}_{i+1}) 
  =
\nonumber \\ && \qquad\qquad = 
  \vec\sigma(\mybasevec + \vec{\alpha}_{j} ) \tau(\mybasevec + \vec{\alpha}_{k}) 
  -
  \tau(\mybasevec + \vec{\alpha}_{j}) \vec\sigma(\mybasevec + \vec{\alpha}_{k} ) 
\\[2mm]  
&&
  \left( \mychi{j} - \mychi{k}\right) 
  \mychis{j}{k} \; 
  \vec\rho(\mybasevec) 
  \tau(\mybasevec + \vec{\alpha}_{i-1} + \vec{\alpha}_{i+1}) 
  = 
\nonumber \\ && \qquad\qquad = 
  \tau(\mybasevec + \vec{\alpha}_{j} ) \vec\rho(\mybasevec + \vec{\alpha}_{k}) 
  -
  \vec\rho(\mybasevec + \vec{\alpha}_{j} ) \tau(\mybasevec + \vec{\alpha}_{k}) 
\\[2mm] 
&&
  \mychis{j}{k} \; 
  \left[ 
    \tau(\mybasevec)
    \tau(\mybasevec + \vec{\alpha}_{j} + \vec{\alpha}_{k}) 
    +
    \left( 
      \vec\rho(\mybasevec), 
      \vec\sigma(\mybasevec + \vec{\alpha}_{j} + \vec{\alpha}_{k}) 
    \right) 
    \right] 
  = 
\nonumber \\ && \qquad\qquad = 
    \tau(\mybasevec + \vec{\alpha}_{j}) 
    \tau(\mybasevec + \vec{\alpha}_{k}) 
\end{eqnarray}
\myendnumparts
with arbitrary $\mychis{j}{k}$ and $\mychi{i}$ satisfying \eref{restr:chi}. 

\end{proposition}

In the following section we consider  the already known bilinear scalar system 
and demonstrate that it can be used to derive solutions for \eref{syst:bilin}. 

\section{Ablowitz-Ladik-Hirota system. \label{sec:hal}}

In this section we discuss the bilinear system, which is closely related to 
\eref{syst:bilin}. After presenting some already known results we derive 
formulae that are used to construct solutions for \eref{syst:bilin} and 
hence for the field equations of our model.

\subsection{Scalar Ablowitz-Ladik-Hirota system.}

Our starting point is the `scalar' version of \eref{syst:bilin}, 
\mynumparts{syst:hal}
\begin{equation}
    \myconst{a}{\alpha,\beta}
    \myvar{\tau} 
    \myShifted{\alpha\beta}{\myvar{\sigma}} 
  = 
    \myShiftedB{\alpha}{\myvar{\sigma}} 
    \myShiftedB{\beta}{\myvar{\tau}} 
  - \myShiftedB{\alpha}{\myvar{\tau}} 
    \myShiftedB{\beta}{\myvar{\sigma}} 
\label{eq:bist}
\end{equation}
\begin{equation}
    \myconst{a}{\alpha,\beta}
    \myvar{\rho} 
    \myShifted{\alpha\beta}{\myvar{\tau}} 
  = 
    \myShiftedB{\alpha}{\myvar{\tau}} 
    \myShiftedB{\beta}{\myvar{\rho}} 
  - \myShiftedB{\alpha}{\myvar{\rho}} 
    \myShiftedB{\beta}{\myvar{\tau}} 
\label{eq:bitr}
\end{equation}
\begin{equation}
    \myvar{\tau} 
    \myShifted{\alpha\beta}{\myvar{\tau}} 
  + \myvar{\rho} 
    \myShifted{\alpha\beta}{\myvar{\sigma}} 
  = 
    \myconst{b}{\alpha\beta}
    \myShiftedB{\alpha}{\myvar{\tau}} 
    \myShiftedB{\beta}{\myvar{\tau}} 
\label{eq:alh}
\end{equation}
\myendnumparts
which we write using the `abstract' shifts $\myShift{\alpha}$. 
Considering the problem of this paper, these shifts should be associated with 
the translations in the auxiliary space $\myauxspace$, 
$ \myShift{\alpha}\!\!: f(\mybasevec) \to f(\mybasevec + \vec\alpha) $, 
and in the final formulae the parameters $\alpha$ and $\beta$ will be taken 
from the set $\{ \alpha_{1}, \alpha_{2}, \alpha_{3} \}$ 
with $\alpha_{i}$ corresponding to the vector $\vec{\alpha}_{i}$ ($i=1,2,3$), 
$
  \left(\myShifted{\alpha_{i}}{f}\right)(\vec{n}) 
  = 
  f(\mybasevec + \vec{\alpha}_{i}) 
$.
However, now we consider $\alpha$ and $\beta$ as arbitrary parameters. 
Moreover, the origin of these shifts and the `inner structure' of the 
tau-functions 
is not important for the time being. 
We are going to study, so to say, algebraic properties of \eref{syst:hal} 
which can be thought of as a system of difference (or functional) 
equations with arbitrary, save the consistency restriction 
\begin{equation}
    \myconst{a}{\alpha,\beta}
    \myconst{b}{\alpha\beta}
  - \myconst{a}{\alpha,\gamma}
    \myconst{b}{\alpha\gamma}
  + \myconst{a}{\beta,\gamma}
    \myconst{b}{\beta\gamma} 
  = 0,
\label{alh-cab}
\end{equation}
skew-symmetric functions $\myconst{a}{\alpha,\beta}$ and symmetric functions 
$\myconst{b}{\alpha\beta}$ of arbitrary parameters $\alpha$ and $\beta$. 

System \eref{syst:hal} is an already known system that appears, in this form 
or another, in studies of a large number of integrable equations. 

It can be shown that an immediate consequence of \eref{eq:bist} and 
\eref{eq:bitr} is the fact that all tau-functions 
($\myvar{\tau}$, $\myvar{\sigma}$ and $\myvar{\rho}$) solve the HBDE:
\begin{equation}
\fl\qquad
  0 
  = 
    \myconst{a}{\alpha,\beta}
    \myShiftedB{\gamma}{\myvar{\omega}} 
    \myShiftedB{\alpha\beta}{\myvar{\omega}} 
  - \myconst{a}{\alpha,\gamma}
    \myShiftedB{\beta}{\myvar{\omega}} 
    \myShiftedB{\alpha\gamma}{\myvar{\omega}} 
  + \myconst{a}{\beta,\gamma}
    \myShiftedB{\alpha}{\myvar{\omega}} 
    \myShiftedB{\beta\gamma}{\myvar{\omega}},
  \quad
  \omega = \tau, \sigma, \rho. 
\label{hal:hbde}
\end{equation}

Another consequence of equations \eref{eq:bist} and \eref{eq:bitr},  
\mynumparts{hal:bt}
\begin{eqnarray}
  0 & = & 
    \myconst{a}{\alpha,\beta}
    \myShiftedB{\gamma}{\myvar{\tau}} 
    \myShiftedB{\alpha\beta}{\myvar{\sigma}} 
  - \myconst{a}{\alpha,\gamma}
    \myShiftedB{\beta}{\myvar{\tau}} 
    \myShiftedB{\alpha\gamma}{\myvar{\sigma}} 
  + \myconst{a}{\beta,\gamma}
    \myShiftedB{\alpha}{\myvar{\tau}} 
    \myShiftedB{\beta\gamma}{\myvar{\sigma}}, 
\label{hal:bt-st}
\\
  0 & = & 
    \myconst{a}{\alpha,\beta}
    \myShiftedB{\gamma}{\myvar{\rho}} 
    \myShiftedB{\alpha\beta}{\myvar{\tau}} 
  - \myconst{a}{\alpha,\gamma}
    \myShiftedB{\beta}{\myvar{\rho}} 
    \myShiftedB{\alpha\gamma}{\myvar{\tau}} 
  + \myconst{a}{\beta,\gamma}
    \myShiftedB{\alpha}{\myvar{\rho}} 
    \myShiftedB{\beta\gamma}{\myvar{\tau}}, 
\label{hal:bt-tr}
\end{eqnarray}
\myendnumparts 
can be interpreted as describing the B\"acklund transformations 
\begin{equation}
  \mbox{BT}_{\mbox{\tiny HBDE}}: \quad 
  \sigma 
  \stackrel{\eref{hal:bt-st}}{\longrightarrow} 
  \tau 
  \stackrel{\eref{hal:bt-tr}}{\longrightarrow}  
  \rho
\end{equation}
between different solutions for the HBDE.

The last of the equations \eref{syst:hal} can be viewed, in the framework 
of the theory of the HBDE, as a nonlinear restriction, which is compatible 
with \eref{eq:bist} and \eref{eq:bitr} (provided the constants  
$\myconst{a}{\alpha,\beta}$ and $\myconst{b}{\alpha\beta}$ meet \eref{alh-cab}).
It turns out that the restricted system \eref{syst:hal} is 
closely related to another integrable model: it describes the action of 
the so-called Miwa shifts of the Ablowitz-Ladik hierarchy (ALH) \cite{AL75}. 

Indeed, the functions 
\begin{equation}
  Q_{n} = \myShift{\kappa}^{n} Q, 
  \qquad
  R_{n} = \myShift{\kappa}^{-n} R 
\end{equation}
where 
\begin{equation}
  \myvar{Q} 
  = 
  \frac{ \myvar{E} }{ \myconst{b}{\kappa\kappa} }
  \frac{ \myShifted{\kappa}{\myvar{\sigma}} }{ \myvar{\tau} }, 
\qquad
  \myvar{R} 
  = 
  \frac{ 1 }{ \myvar{E} } 
  \frac{ \myShift{\kappa}^{-1} \myvar{\rho} }{ \myvar{\tau} }
\end{equation}
and $\myvar{E}$ (the discrete analogue of the plane-wave background) 
is defined by 
$
  \myShifted{\alpha}{\myvar{E}} 
  = 
  \myvar{E} / \myconst{b}{\alpha\kappa} 
$
satisfy, for a fixed value of $\kappa$,
\mynumparts{}
\begin{eqnarray}
  \myShift[E]{\alpha} Q_{n} - Q_{n} 
  & = & 
  \xi_{\alpha} 
  \left[ 1 - R_{n} \left( \myShift[E]{\alpha} Q_{n} \right) \right]
  \myShift[E]{\alpha} Q_{n+1}, 
\\
  R_{n} - \myShift[E]{\alpha} R_{n}  
  & = & 
  \xi_{\alpha} 
  \left[ 1 - R_{n} \left( \myShift[E]{\alpha} Q_{n} \right) \right]
  R_{n-1}. 
\end{eqnarray}
\myendnumparts 
where 
$\myShift[E]{\alpha} = \myShift{\alpha}\myShift{\kappa}^{-1}$ 
and 
$\xi_{\alpha} =  \myconst{a}{\alpha,\kappa}\myconst{b}{\alpha\kappa}$.
These equations, 
with $\myShift[E]{\alpha}$ being interpreted as the Miwa shift with respect to 
$\xi_{\alpha}$, 
are nothing but the so-called functional representation of the 
positive flows of the ALH \cite{V98,V02}. 

What is important for our present study is that the HBDE and the ALH 
(and hence the system \eref{syst:hal}) are integrable models, which during 
their 40-year history have attracted considerable interest and which are one 
of the very well studied integrable systems. 
Thus, one can use various results that have been obtained for the HBDE 
and the ALH to derive solutions for \eref{syst:hal} and hence for system 
\eref{syst:bilin}.

\subsection{Vector Ablowitz-Ladik-Hirota system.}

Now we demonstrate how to construct, starting from solutions for 
\eref{syst:hal}, solutions for another system, which is a vector 
generalization of \eref{syst:hal}. 

To do this, we first derive two Miura-like transformations 
$
  \left( \tau, \sigma, \rho \right)
  \to 
  \left( \tau, \sigma^{(m)}_{\kappa}, \rho^{(m)}_{\kappa} \right)
$, 
($m=2,3$)  
which then can be used to form the scalar-vector triplet 
$ \left( \tau, \vec\sigma, \vec\rho \right) $.

The key result of this section can be formulated as 
%
\begin{proposition} \label{prop:kappa}
If $\tau$, $\sigma$ and $\rho$ solve equations \eref{syst:hal} 
then the new tau-functions $\sigma^{(m)}_{\kappa}$ and $\rho^{(m)}_{\kappa}$ 
($m=2,3$) given by 
\begin{equation}
  \myvar{\mytauQ_{\kappa}} 
  = 
  v_{\kappa}^{-1} \myShifted{\kappa}{\myvar{\sigma}}, 
  \qquad
  \myvar{\mytauR_{\kappa}} 
  = 
  v_{\kappa}  \myShiftedInv{\kappa}{\myvar{\rho}} 
\label{def:sr-2}
\end{equation}
and 
\begin{equation}
  \myvar{\mytauV_{\kappa}} 
  = 
  u_{\kappa}^{-1}  \myShiftedInv{\kappa}{\myvar{\tau}},
  \qquad 
  \myvar{\mytauU_{\kappa}} 
  = 
  u_{\kappa} \myShifted{\kappa}{\myvar{\tau}} 
\label{def:sr-3}
\end{equation}
with functions $u_{\kappa}$ and $v_{\kappa}$ defined by 
\begin{equation}
	\myShifted{\xi}{u_{\kappa}} = \myconst{a}{\kappa,\xi} u_{\kappa},
  \qquad
  \myShifted{\xi}{v_{\kappa}} = \myconst{b}{\kappa\xi} v_{\kappa}
\label{kappa:uv}
\end{equation}
solve \eref{eq:bist} and \eref{eq:bitr},
\mynumparts{syst:kappa}
\begin{eqnarray}
    \myconst{a}{\alpha,\beta}
    \mathop{\myvar{\tau}} 
    \myShifted{\alpha\beta}{\myvar{\sigma^{(2,3)}_{\kappa}}} 
  & = & 
    \myShiftedB{\alpha}{\myvar{\sigma^{(2,3)}_{\kappa}}} 
    \myShiftedB{\beta}{\myvar{\tau}} 
  - \myShiftedB{\alpha}{\myvar{\tau}} 
    \myShiftedB{\beta}{\myvar{\sigma^{(2,3)}_{\kappa}}}, 
\label{eq:kappa-s}
\\
    \myconst{a}{\alpha,\beta} 
    \myvar{\rho^{(2,3)}_{\kappa}} 
    \myShifted{\alpha\beta}{\myvar{\tau}} 
  & = & 
    \myShiftedB{\alpha}{\myvar{\tau}} 
    \myShiftedB{\beta}{\myvar{\rho^{(2,3)}_{\kappa}}} 
  - \myShiftedB{\alpha}{\myvar{\rho^{(2,3)}_{\kappa}}} 
    \myShiftedB{\beta}{\myvar{\tau}}, 
\label{eq:kappa-r}
\end{eqnarray}
and are related by 
\begin{equation}
  \myvar{\mytauR_{\kappa}} 
  \myShifted{\alpha\beta}{\myvar{\mytauQ_{\kappa}}} 
  + 
  \myvar{\mytauU_{\kappa}} 
  \myShifted{\alpha\beta}{\myvar{\mytauV_{\kappa}}} 
  = 
  \myconst{\hat{b}}{\alpha\beta,\kappa}
  \myShiftedB{\alpha}{\myvar{\tau}} 
  \myShiftedB{\beta}{\myvar{\tau}} 
\label{eq:kappa-t}
\end{equation}
\myendnumparts
where 
\begin{equation}
 \myconst{\hat{b}}{\alpha\beta,\kappa} 
  = 
  \frac{ 1 }{ \myconst{a}{\alpha,\kappa} \myconst{a}{\beta,\kappa} }
  \frac{ \myconst{b}{\alpha\beta}  \myconst{b}{\kappa\kappa} }
       { \myconst{b}{\alpha\kappa} \myconst{b}{\beta\kappa}  }.
\label{def:hatb}
\end{equation}
\end{proposition}
(see appendix for a proof).

Now, one can easily obtain solutions for the vector generalization of 
\eref{syst:hal}: vectors $\vec{\sigma}$ and $\vec\rho$, 
\mynumparts{def:vsr}
\begin{eqnarray}
  \vec{\sigma} 
  & = & 
  \left( 
    \myvar{\sigma}, 
    \myvar{\mytauQ_{\kappa}}, 
    \myvar{\mytauV_{\kappa}}  
  \right)^{T} 
  \in \mathbb{R}^{3}
  \\[2mm]
  \vec{\rho} 
  & = & 
  \left( 
    \myvar{\rho}, 
    \myvar{\mytauR_{\kappa}}, 
    \myvar{\mytauU_{\kappa}} 
  \right)^{T}
  \in \mathbb{R}^{3}
\end{eqnarray}
\myendnumparts
(we do not indicate the dependence on $\kappa$ in the left hand side of 
\eref{def:vsr} considering it as a fixed parameter) satisfy the vector 
variant of \eref{eq:bist}--\eref{eq:bitr},  
\mynumparts{syst:halv}
\begin{equation}
    \myconst{a}{\alpha,\beta}
    \myvar{\tau} 
    \myShifted{\alpha\beta}{\vec\sigma} 
  = 
    \myShiftedB{\alpha}{\vec\sigma} 
    \myShiftedB{\beta}{\myvar{\tau}} 
  - \myShiftedB{\alpha}{\myvar{\tau}} 
    \myShiftedB{\beta}{\vec\sigma} 
\end{equation}
\begin{equation}
    \myconst{a}{\alpha,\beta}
    \vec\rho 
    \myShifted{\alpha\beta}{\myvar{\tau}} 
  = 
    \myShiftedB{\alpha}{\myvar{\tau}} 
    \myShiftedB{\beta}{\vec\rho} 
  - \myShiftedB{\alpha}{\vec\rho} 
    \myShiftedB{\beta}{\myvar{\tau}} 
\end{equation}
and are related by the vector variant of \eref{eq:alh},
\begin{equation}
    \myvar{\tau} 
    \myShifted{\alpha\beta}{\myvar{\tau}} 
  + \left( 
      \vec\rho, 
      \myShifted{\alpha\beta}{\vec\sigma} 
    \right)
  = 
    \myconst{c}{\alpha\beta}
    \myShiftedB{\alpha}{\myvar{\tau}} 
    \myShiftedB{\beta}{\myvar{\tau}} 
\end{equation} 
\myendnumparts
with 
\begin{equation}
  \myconst{c}{\alpha\beta} 
  = 
  \myconst{b}{\alpha\beta} + \myconst{\hat{b}}{\alpha\beta,\kappa}. 
\label{def:cbb}
\end{equation}

\subsection{Solutions for \eref{syst:bilin}.}

The system \eref{syst:halv} is, up to the constants, nothing but the bilinear 
system \eref{syst:bilin}. To make them coincide, one has 
i) to identify the translations in the auxiliary space $\myauxspace$, 
$f(\mybasevec) \to f(\mybasevec + \vec\alpha_{i}) $ with 
action of $\myShift{\alpha_{i}}$ ($i=1,2,3$), 
where 
$\left\{ \alpha_{i} \right\}_{i=1,2,3}$ together with $\kappa$ is a set of 
parameters, describing solution, 
ii) to note that $ \mychis{j}{k} = 1 / \myconst{c}{\alpha_{j}\alpha_{k}} $
and 
iii) to ensure \eref{restr:chi} for the quantities $\mychi{i}$ 
that should be found from 
\begin{equation}
  \mychi{j} - \mychi{k} 
  = 
  \myconst{a}{\alpha_{j},\alpha_{k}} 
  \myconst{c}{\alpha_{j},\alpha_{k}}, 
\label{eq:chi}
\end{equation}
which leads to some restrictions on $\alpha_{i}$ and $\kappa$. 
However, we do not solve this problem now and return to it later 
because 
from the practical viewpoint, the application of the presented results is 
as follows: 
\begin{itemize}
	\item 
	we select the class of solutions we want to obtain 
	(for example, soliton, finite-gap or Toeplitz), 
	\item
	we take a set of identities 
	(for example, the Fay identities or the Jacobi determinant identities) 
	for the objects that are used to construct these solutions 
	(determinants of the Cauchy-like matrices, the theta-functions 
	or the 	Toeplitz determinants) 
	and present them in form similar to \eref{syst:hal}, 
	\item
	knowing 
	$\myconst{a}{\alpha,\beta}$ and $\myconst{b}{\alpha\beta}$, 
	which depend on the identities we use, 
	we establish the relationships between the 	parameters 
	(in our case $\alpha_{i}$, $\kappa$ and $\Gamma_{i}$), 
	\item
	we use the above formulae to construct 
	$\vec\sigma$ and $\vec\rho$ and then $\vec{\phi}$. 
\end{itemize}

In the next section we employ this algorithm to derive the $N$-soliton 
solutions for our model.

\section{$N$-soliton solutions. \label{sec:sls}}

To derive the $N$-soliton solutions for our model we use the results of 
\cite{V15} where we have presented a large number of the so-called soliton Fay 
identities for the $N{\times}N$ matrices of a special type, which solve the 
Sylvester equation 
\begin{equation}
  \begin{array}{lcl}
  \mymatrix{L} \mymatrix{A} - \mymatrix{A} \mymatrix{R} 
  & = & 
  \myunitket \mybra{a}, 
  \\
  \mymatrix{R} \mymatrix{B} - \mymatrix{B} \mymatrix{L} 
  & = & 
  \myunitket \mybra{b} 
  \end{array}
\label{eq:Sylvester} 
\end{equation}
where 
$\mymatrix{L}$ and $\mymatrix{R}$ are diagonal constant $N{\times}N$ matrices, 
\begin{equation}
  \begin{array}{lcl}	
  \mymatrix{L} & = & \mbox{diag}\left( L_{1}, ... , L_{N} \right), 
  \\
  \mymatrix{R} & = & \mbox{diag}\left( R_{1}, ... , R_{N} \right), 
  \end{array}
\label{sls:defLR} 
\end{equation}
$\myunitket$ is the $N$-column with all components equal to $1$ 
(note that we have replaced the $N$-columns 
$| \,\alpha\, \rangle$ and $| \,\beta\, \rangle$ used in \cite{V15} 
with  $\myunitket$, which can be done by means of the simple gauge transform), 
$\mybra{a}$ and $\mybra{b}$ are $N$-component rows that depend on the 
coordinates describing the model. 

The shifts $\myShift{\zeta}$ are defined by 
\begin{equation}
  \begin{array}{lcl}
  \myShifted{\zeta}{\mybra{a}} 
  & = & 
  \mybra{a} \left(\mymatrix{R} - \zeta\right)^{-1}, 
  \\[2mm]
  \myShifted{\zeta}{\mybra{b}} 
  & = & 
  \mybra{b} \left(\mymatrix{L} - \zeta\right) 
  \end{array} 
\label{sls:shifts}
\end{equation}
which determines the shifts of all other objects 
(the matrices $\mymatrix{A}$ and $\mymatrix{B}$, their determinants, 
the tau-functions constructed of $\mymatrix{A}$ and $\mymatrix{B}$ etc).

The soliton tau-functions have been defined in \cite{V15} as  
\mynumparts{sls:tsr}
\begin{equation}
  \tau 
  = 
  \det \left| \mymatrix{1} + \mymatrix{A}\mymatrix{B} \right|
\end{equation}
and 
\begin{eqnarray}
  \sigma & = & \tau \langle a | \mymatrix{F} \myunitket, 
  \\ 
  \rho   & = & \tau \langle b | \mymatrix{G} \myunitket 
\end{eqnarray} 
\myendnumparts
where matrices $\mymatrix{F}$ and $\mymatrix{G}$ are given by 
\mynumparts{sls:FG}
\begin{eqnarray} 
  \mymatrix{F} 
  & = & 
  ( \mymatrix{1} + \mymatrix{B}\mymatrix{A} )^{-1}, 
  \\ 
  \mymatrix{G} 
  & = & 
  ( \mymatrix{1} + \mymatrix{A}\mymatrix{B} )^{-1}. 
\end{eqnarray} 
\myendnumparts
The simplest soliton Fay identities, which are equations (3.12)--(3.14) 
of \cite{V15}, 
are exactly equations \eref{eq:alh}, \eref{eq:bist} and \eref{eq:bitr} with  
\begin{equation}
  \myconst{a}{\alpha,\beta} = \alpha - \beta, 
  \qquad
  \myconst{b}{\alpha\beta} = 1 
\label{sls:ab}
\end{equation}
Thus, 
\eref{eq:Sylvester}--\eref{sls:FG} provide solutions for \eref{syst:hal} which, 
by means of the recipe of proposition \ref{prop:kappa}, 
yield the vector tau-functions \eref{def:vsr}. 
To simplify the final formulae, one can use the matrix identities derived in 
\cite{V15} (see equations (2.9)--(2.12) of \cite{V15}),  
\mynumparts{}
\begin{eqnarray} 
  \myShiftedB{\zeta}{\tau} / \tau 
  & = & 
  1 - 
  \langle b | \mymatrix{G} \, \myShiftedB{\zeta}{\mymatrix{A}} \myunitket, 
\\ 
  \myShiftedInvB{\zeta}{\tau} / \tau 
  & = & 
  1 - 
  \langle a | \mymatrix{F} \, \myShiftedInvB{\zeta}{\mymatrix{B}} \myunitket
\end{eqnarray} 
\myendnumparts 
and
\mynumparts{sls:tt}
\begin{eqnarray}
  \myShiftedB{\zeta}{\sigma} / \tau 
  & = & 
  \langle a | \mymatrix{F} \, (\mymatrix{R} - \zeta)^{-1} \myunitket, 
\\
  \myShiftedInvB{\zeta}{\rho} / \tau 
  & = & 
  \langle b | \mymatrix{G} \, (\mymatrix{L} - \zeta)^{-1} \myunitket. 
\end{eqnarray}
\myendnumparts 

The only thing we have to do to derive the soliton solutions 
is to settle the question of the parameters. To this end we have 
to express, using \eref{eq:chi}, $\mychi{i}$ in terms of $\alpha_{i}$ 
and to ensure \eref{restr:chi}.

From \eref{def:cbb}, \eref{def:hatb} and \eref{sls:ab} one can get 
\begin{equation}
  \myconst{c}{\alpha,\beta} 
  = 
  1 
  + 
  \frac{ 1 }{ (\alpha - \kappa)(\beta - \kappa) } 
\end{equation}
which yields, together with \eref{eq:chi}, 
\begin{equation}
  \mychi{i} 
  = 
  \alpha_{i} 
  - 
  \frac{ 1 }{ \alpha_{i} - \kappa }. 
\end{equation}
It is easy to see that one can meet \eref{restr:chi} without imposing any 
restrictions on $\alpha_{i}$ ($i=1,2,3$) by choosing 
$\kappa=\kappa\left( \{ \alpha_{i} \}, \{ \Gamma_{i} \} \right)$ 
as a solution of the equation
\begin{equation}
  \sum_{i=1}^{3} 
  \Gamma_{i} 
  \left( \alpha_{i} - \frac{ 1 }{ \alpha_{i} - \kappa } \right) 
  = 0 
\label{sls:kappa}
\end{equation}
which can be rewritten as a cubic one,
\begin{equation}
  \prod_{i=1}^{3}  \left( \kappa - \alpha_{i} \right) 
  + \kappa 
  + C\left( \{ \alpha_{i} \}, \{ \Gamma_{i} \} \right) = 0
\label{sls:cubic}
\end{equation}
with 
$
  C\left( \{ \alpha_{i} \}, \{ \Gamma_{i} \} \right) = 
  \left. \sum_{i=1}^{3} \Gamma_{i} \alpha_{i-1}\alpha_{i+1} \right/ 
       \sum_{i=1}^{3} \Gamma_{i} \alpha_{i} 
$.

Thus, we have all necessary to write down the $N$-soliton solutions.
The results of section \ref{sec:hal} together with 
\eref{sls:tsr}--\eref{sls:tt} give us the structure of solutions, 
\mynumparts{}
\begin{eqnarray}
  \myphip 
  & = & 
  \frac{1}{u} \, 
  ( 0, 0, 1 )^{\scriptscriptstyle T}
  +
  \sum_{\ell=1}^{N} f_{\ell} \, \myPhip_{\ell}, 
  \\ 
  \myphin 
  & = & 
  u \, ( 0, 0, 1 )^{\scriptscriptstyle T}
  +
  \sum_{\ell=1}^{N} g_{\ell} \, \myPhin_{\ell}
\end{eqnarray}
\myendnumparts
where we write $u$ instead of $u_{\kappa}$ and put $v_{\kappa}=1$ 
(which follows from \eref{kappa:uv} and \eref{sls:ab}),
\mynumparts{}
\begin{eqnarray}
  \myPhip_{\ell} 
  & = & 
  \left( 
    1 , \;  
    \frac{1}{R_{\ell}-\kappa}, \; 
    - \frac{1}{u} \sum_{m=1}^{N} \mymatrix{B}_{\ell m} 
      \frac{1}{L_{m}-\kappa} 
  \right)^{\scriptscriptstyle T} 
  \\ 
  \myPhin_{\ell} 
  & = & 
  \left( 
    1 , \;  
    \frac{1}{L_{\ell}-\kappa}, \; 
    - u \sum_{m=1}^{N} \mymatrix{A}_{\ell m} 
      \frac{1}{R_{m}-\kappa} 
  \right)^{\scriptscriptstyle T} 
\end{eqnarray}
\myendnumparts
and $f_{\ell}$ and $g_{\ell}$ are components of the $N$-rows 
$\mybra{a} \mymatrix{F}$ and $\mybra{b} \mymatrix{G}$, 
\mynumparts{sls:fg}
\begin{eqnarray}
  \left( f_{1}, ..., f_{N} \right) 
  & = & 
  \mybra{a} \mymatrix{F}, 
  \\ 
  \left( g_{1}, ..., g_{N} \right) 
  & = & 
  \mybra{b} \mymatrix{G}. 
\end{eqnarray}
\myendnumparts
The dependence of $\myphip$ and $\myphin$ and hence of $\vec\phi$ on the 
coordinates (i.e. on $\vec{n}$) is given by \eref{def:nu} and the 
correspondence $\vec{\alpha}_{i} \to \myShift{\alpha_{i}}$. Now, we want to 
eliminate the auxiliary vectors $\vec{\alpha}_{i}$ and present the soliton 
solutions as functions of $\vec{n}$. To this end, we rewrite \eref{def:qr} 
as 
\begin{equation}
  \vec{\phi}(\vec{n}) 
  = 
  \left\{ 
  \begin{array}{rcl} 
    \prod_{i=1}^{3} 
    \myShift{\alpha_{i}}^{ n_{i} - \mathcal{N}( n_{1},n_{2},n_{3}) } 
    \myphip 
    && (\myin{+})
    \\[2mm] 
    \prod_{i=1}^{3} 
    \myShift{\alpha_{i}}^{ n_{i} - \mathcal{N}( n_{1},n_{2},n_{3}) } 
    \myphin 
    && (\myin{-}) 
  \end{array}
  \right.
\end{equation}
where 
\begin{equation}
  \mathcal{N}(n_{1},n_{2},n_{3}) 
  = 
  \left\{ 
  \begin{array}{lcl} 
    \frac{1}{3} 
    \sum_{i=1}^{3} n_{i} 
    && (\myin{+}) 
    \\[2mm] 
    \frac{1}{3} 
    \left( \sum_{i=1}^{3} n_{i} + 2 \right) 
    && (\myin{-}) 
  \end{array}
  \right.
\end{equation}
(note that $\mathcal{N}(n_{1},n_{2},n_{3})$ is integer for any 
$\vec{n} \in \Lambda^{\pm}$, and that differences 
$n_{i} - \mathcal{N}(n_{1},n_{2},n_{3})$ are invariant under the simultaneous 
shift $n_{i} \to n_{i} + 1$, $i=1,2,3$).
Thus, one can express the dependence of soliton tau-functions on $\vec{n}$ 
by introducing the diagonal matrices
\mynumparts{sls:LRn}
\begin{eqnarray}
  \mymatrix{L}(\vec{n}) 
  & = & 
  \prod_{i=1}^{3} 
  \left( 
    \mymatrix{L} - \alpha_{i} 
  \right)^{ n_{i} - \mathcal{N}( n_{1},n_{2},n_{3}) }, 
  \\ 
  \mymatrix{R}(\vec{n}) 
  & = & 
  \prod_{i=1}^{3} 
  \left( 
    \mymatrix{R} - \alpha_{i} 
  \right)^{ n_{i} - \mathcal{N}( n_{1},n_{2},n_{3}) }. 
\end{eqnarray}
\myendnumparts
The definitions of the shifts \eref{sls:shifts} lead to 
\begin{equation}
  \mybra{a(\vec{n})} 
  = 
  \mybra{a_{0}} \mymatrix{R}(\vec{n})^{-1}, 
  \qquad 
  \mybra{b(\vec{n})} 
  = 
  \mybra{b_{0}} \mymatrix{L}(\vec{n}) 
\end{equation}
where $\mybra{a_{0}}$ and $\mybra{b_{0}}$ are constant $N$-rows 
and similar formulae for $\mymatrix{A}(\vec{n})$ and 
$\mymatrix{B}(\vec{n})$:
\begin{equation}
  \mymatrix{A}(\vec{n}) 
  = 
  \mymatrix{A}_{0} \mymatrix{R}(\vec{n})^{-1}, 
  \qquad 
  \mymatrix{B}(\vec{n}) 
  = 
  \mymatrix{B}_{0} \mymatrix{L}(\vec{n}) 
\label{sls:ABn}
\end{equation}
with constant $\mymatrix{A}_{0}$ and $\mymatrix{B}_{0}$ 
(which are, recall, related to $\mybra{a_{0}}$ and $\mybra{b_{0}}$ 
by \eref{eq:Sylvester}). 
The definitions \eref{sls:fg} of $f_{\ell}$ and $g_{\ell}$ can be 
rewritten as 
\mynumparts{sls:fgn}
\begin{eqnarray}
  f_{\ell}(\vec{n}) 
  & = & 
  \sum_{m=1}^{N} 
  \left( \mymatrix{K} \mymatrix{X}(\vec{n}) \right)_{m\ell}, 
\\ 
  g_{\ell}(\vec{n}) 
  & = & 
  - \sum_{m=1}^{N} 
    \left( \mymatrix{K}^{\scriptscriptstyle T} \mymatrix{Y}(\vec{n}) \right)_{m\ell}  
\end{eqnarray}
\myendnumparts
where 
$(...)_{ml}$ denotes the element of a $N{\times}N$ matrix, 
$\mymatrix{K}$ is the inverse of the matrix with the elements 
$1/(L_{l} - R_{m})$, 
\begin{equation}
  \mymatrix{K} = \tilde\mymatrix{K}^{-1}, 
  \qquad
  \tilde\mymatrix{K} 
  = 
  \left( \frac{ 1 }{ L_{l} - R_{m} } \right)_{l,m=1,...,N}  
\end{equation}
and
\mynumparts{sls:XY}
\begin{eqnarray}
  \mymatrix{X}(\vec{n}) 
  & = & 
  \left[ \mymatrix{A}^{-1}(\vec{n}) + \mymatrix{B}(\vec{n}) \right]^{-1}, 
\\ 
  \mymatrix{Y}(\vec{n}) 
  & = & 
  \left[ \mymatrix{A}(\vec{n}) + \mymatrix{B}^{-1}(\vec{n}) \right]^{-1}. 
\end{eqnarray}
\myendnumparts
Finally, an analysis of \eref{kappa:uv} and \eref{sls:ab}, together with 
\eref{def:qr} leads to 
\begin{equation}
  u(\vec{n}) 
  = 
  \prod_{i=1}^{3} 
  \left( 
    \kappa - \alpha_{i} 
  \right)^{ n_{i} - \mathcal{N}( n_{1},n_{2},n_{3}) }, 
\label{sls:u}
\end{equation}
To summarize, we can present the main result of this paper as 
%
\begin{proposition}\label{prop:sls}
The $N$-soliton solutions for the field equations \eref{syst:main} 
are given by 
\begin{equation}
  \vec{\phi}(\vec{n}) 
  = 
  \left\{ 
  \begin{array}{rcl} 
  \displaystyle 
  \frac{ 1 }{ u(\vec{n}) } \, \vec{\phi}_{*} 
  +
  \sum_{\ell=1}^{N} f_{\ell}(\vec{n}) \myPhip_{\ell}(\vec{n})
  && ( \myin{+} ) 
  \\  
  \displaystyle
  u(\vec{n}) \, \vec{\phi}_{*} 
  +
  \sum_{\ell=1}^{N} g_{\ell}(\vec{n}) \myPhin_{\ell}(\vec{n})
  && ( \myin{-} ) 
  \end{array}
  \right.
\end{equation}
where $\vec{\phi}_{*} = ( 0, 0, 1 )^{\scriptscriptstyle T}$, 
the scalars $u(\vec{n})$, $f_{\ell}(\vec{n})$ and $g_{\ell}(\vec{n})$ 
are given by \eref{sls:u} and \eref{sls:fgn}--\eref{sls:XY}, 
the vectors $\myPhip_{\ell}(\vec{n})$ and $\myPhin_{\ell}(\vec{n})$ 
are given by 
\mynumparts{}
\begin{eqnarray}
  \myPhip_{\ell}(\vec{n}) 
  & = & 
  \left( 
    1 , \;  
    \frac{1}{R_{\ell}-\kappa}, \; 
    - \frac{1}{u(\vec{n})} \sum_{m=1}^{N} \mymatrix{B}_{\ell m}(\vec{n}) 
      \frac{1}{L_{m}-\kappa} 
  \right)^{\scriptscriptstyle T}, 
  \\ 
  \myPhin_{\ell}(\vec{n}) 
  & = & 
  \left( 
    1 , \;  
    \frac{1}{L_{\ell}-\kappa}, \; 
    - u(\vec{n}) \sum_{m=1}^{N} \mymatrix{A}_{\ell m}(\vec{n}) 
      \frac{1}{R_{m}-\kappa} 
  \right)^{\scriptscriptstyle T}, 
\end{eqnarray}
\myendnumparts
the matrices $\mymatrix{A}(\vec{n})$ and $\mymatrix{B}(\vec{n})$ 
are defined in \eref{sls:ABn} and \eref{sls:LRn} with  
\begin{equation}
\fl\qquad
  \mymatrix{A}_{0} 
  = 
  \left( \frac{ a_{0m} }{ L_{l} - R_{m} } \right)_{l,m=1,...,N}, 
  \qquad
  \mymatrix{B}_{0} 
  = 
  \left( \frac{ b_{0m} }{ R_{l} - L_{m} } \right)_{l,m=1,...,N}. 
\end{equation} 
Here, $u_{0}$, $L_{n}$, $R_{n}$, $a_{0n}$, $b_{0n}$ ($n=1,...,N$) 
and $\alpha_{i}$ ($i=1,2,3$) are arbitrary constants and 
$ \kappa = \kappa( \{\alpha_{i}\},\{\Gamma_{i}\}) $ 
is defined in \eref{sls:kappa}.

\end{proposition}
 
\subsection*{$1$-soliton solution.}

To illustrate the obtained results, let us write down the $1$-soliton solution. 
Of course, all we need is just to simplify the formulae of the Proposition 
\ref{prop:sls} taking into account that in the $N=1$ case all matrices become 
scalars: $\mymatrix{L}=L$, $\mymatrix{R}=R$ etc 
(we omit the index $1$ in the definitions \eref{sls:defLR}). 
However, we use some simple transformations to present these solutions in the 
$\exp(...)/\cosh(...)$ form, which is usual for the physical literature.

Hereafter, we take $\vec{e}_{i}$ to be 
unit vectors, with $2\pi/3$ angle between the different ones,
\begin{equation}
  \left( \vec{e}_{i}, \vec{e}_{i} \right) = 1, 
  \quad
  \left( \vec{e}_{i}, \vec{e}_{i \pm 1} \right) = -1/2 
  \qquad
  i=1,2,3. 
\end{equation}
Now, note that the typical product describing the 
$\vec{n}$-dependence, for example in \eref{sls:LRn}, can be written as 
\begin{equation}
  \prod_{i=1}^{3} 
  \left( x - \alpha_{i} \right)^{ n_{i} - \frac{1}{3}\sum_{k=1}^{3} n_{k} } 
  = 
  \exp \left( \myslambda(x), \vec{n} \right) 
\end{equation}
where 
\begin{equation}
  \myslambda(x) 
  = 
  \frac{2}{3} 
  \sum_{i=1}^{3} 
    \ln\left( x - \alpha_{i} \right) \vec{e}_{i}. 
\label{sls:lambda}
\end{equation}
Thus, one can rewrite, for example, the definition of $\mymatrix{L}(\vec{n})$ 
as 
\begin{equation}
  \mymatrix{L}(\vec{n}) 
  = 
  \mysmu(L)^{\pm 1} 
  \exp \left( \myslambda(L), \vec{n} \right) 
  \qquad 
  (\myin{\pm}) 
\end{equation}
with
\begin{equation}
  \mysmu(x) 
  = 
  \left[ 
  \prod\nolimits_{i=1}^{3} 
    \left( x - \alpha_{i} \right) 
  \right]^{1/3}. 
\label{sls:mu}
\end{equation}
After presenting the matrices $\mymatrix{A}(\vec{n})$ and 
$\mymatrix{B}(\vec{n})$ in the similar way and substituting them into 
the general formula, one can obtain the following expression for the 
$1$-soliton solution:
\begin{equation} 
\fl\qquad
  \vec{\phi}(\vec{n}) 
  = 
  \frac{ 1 }{ \cosh\left[ \mysphase{0} \pm \mysconst{0} \right] }
  \left( 
  \begin{array}{l} 
  \pm c_{1} \exp\left[ \pm \mysphase{1} \right] 
  \\
  \pm c_{2} \exp\left[ \pm \left(\mysphase{1}+\Delta\right) \right] 
  \\ 
  c_{3} 
  \exp\left[ \pm \mysphase{2} \right] 
  \cosh\left[ \mysphase{0} \pm \mysconst{1} \right] 
  \end{array} 
  \right)
  \quad
  (\myin{\pm}). 
\label{sls:one}
\end{equation}  
Here, the new functions $\mysphase{1,2,3}$ are introduced by 
$
  \mymatrix{A}(\vec{n}) 
  \propto 
  e^{\mysphase{0} + \mysphase{1} } 
$, 
$
  \mymatrix{B}(\vec{n}) 
  \propto 
  e^{\mysphase{0} - \mysphase{1} } 
$,
$
  u(\vec{n}) 
  \propto 
  e^{ - \mysphase{2} } 
$
and can be written as 
\mynumparts{}
\begin{eqnarray}
  \mysphase{0} 
  & = & 
  \phantom{-} 
  \myhalf 
  \left( \, \myslambda(L) - \myslambda(R), \vec{n} \, \right), 
\\ 
  \mysphase{1} 
  & = & 
  - 
  \myhalf 
  \left( \, \myslambda(L) + \myslambda(R), \vec{n} \, \right), 
\\ 
  \mysphase{2} 
  & = & 
  - 
  \left( \, \myslambda(\kappa) , \vec{n} \, \right). 
\end{eqnarray}
\myendnumparts 
The new constants $\mysconst{0,1}$, $\Delta$ and $c_{1,2,3}$ are given by 
\mynumparts{}
\begin{eqnarray}
  \mysconst{0} 
  & = & 
  \myhalf 
  \left[ \ln\mysmu(L) - \ln\mysmu(R) \right], 
\\ 
  \Delta 
  & = & 
  \myhalf 
  \left[ \ln(L - \kappa) - \ln(R - \kappa) \right], 
\\ 
  \mysconst{1} 
  & = & 
  \mysconst{0} - \Delta 
\end{eqnarray}
\myendnumparts 
and 
\mynumparts{}
\begin{eqnarray}
  c_{1} & = & \myhalf \left(L - R\right) \mysmu(L)^{-1/2} \mysmu(R)^{-1/2}, 
\\ 
  c_{2} & = & (L - \kappa)^{-1/2} (R - \kappa)^{-1/2} \; c_{1},  
\\ 
  c_{3} 
  & = & 
  \mysmu(\kappa)^{-1}. 
\end{eqnarray}
\myendnumparts 
Clearly, these formulae are valid (produce real solutions) only when 
$L,R > \alpha_{1},\alpha_{2},\alpha_{3},\kappa$ or 
$L,R < \alpha_{1},\alpha_{2},\alpha_{3},\kappa$. 
If these inequalities do not hold, then one has to rewrite 
\eref{sls:lambda} and \eref{sls:mu} replacing 
$\ln\left( x - \alpha_{i} \right)$ with
$\ln\left| x - \alpha_{i} \right|$ 
which results in different distributions of the $\pm$ signs in front of the 
constants $c_{1,2,3}$ in \eref{sls:one}. 
In fact, the complete analysis even of the one-soliton solution is rather 
tedious: one has to enumerate all possible positions of $L$ and $R$ with 
respect to $\alpha_{1},\alpha_{2},\alpha_{3}$ as well as all possible choices 
of $\kappa$ as a root of the \emph{cubic} equations \eref{sls:cubic}. 
This leads to different `soliton branches' and in some cases to the singular 
solutions 
(when $\cosh\left[ \mysphase{0} \pm \mysconst{0} \right]$ 
becomes $\sinh\left[ \mysphase{0} \pm \mysconst{0} \right]$) .

\section{Conclusion.}

To conclude we would like to enumerate the main points of our derivation of 
solutions for the problem considered in this paper.

The first step is the substitution \eref{def:nu} which enables to rewrite 
the field equations \eref{syst:main}, 
different for $\Lambda^{+}$ and $\Lambda^{-}$, 
as \emph{one} translationally-invariant 
system \eref{syst:qr}. 
The second part is the ansatz \eref{def:ansatz} which leads to the 
\emph{bilinear} system \eref{syst:bilin}.
Next, the results of proposition \ref{prop:kappa} 
which reduce the \emph{vector} equations \eref{syst:bilin} 
to the \emph{scalar} system \eref{syst:hal}.
Finally, we identified \eref{syst:hal} with the well-studied integrable 
systems which enabled to apply already known results to our problem. 

We hope that similar procedure can lead to solution of other problems 
on non-square lattices.

\appendix
\renewcommand\mythesection{\Alph{section}}

\section{Proof of Proposition \ref{prop:kappa}. \label{app-kappa}}

To prove \eref{eq:kappa-s} and \eref{eq:kappa-r} for \eref{def:sr-3} 
is rather easy. 
Equation \eref{hal:hbde} with $\omega=\tau$ and $\gamma=\kappa$ 
\begin{equation}
  0
  = 
    \myconst{a}{\beta,\kappa}
    \myShiftedB{\alpha}{\myvar{\tau}} 
    \myShiftedB{\beta\kappa}{\myvar{\tau}} 
  - \myconst{a}{\alpha,\kappa}
    \myShiftedB{\beta}{\myvar{\tau}} 
    \myShiftedB{\alpha\kappa}{\myvar{\tau}} 
  + \myconst{a}{\alpha,\beta}
    \myShiftedB{\kappa}{\myvar{\tau}} 
    \myShiftedB{\alpha\beta}{\myvar{\tau}} 
\end{equation}
and the last equation after application of $\myShift{\kappa}^{-1}$,
\begin{equation}
  0
  = 
    \myconst{a}{\beta,\kappa}
    \myShiftedB{\beta}{\myvar{\tau}} 
    \myShiftedB{\alpha\bar{\kappa}}{\myvar{\tau}} 
  - \myconst{a}{\alpha,\kappa}
    \myShiftedB{\alpha}{\myvar{\tau}} 
    \myShiftedB{\beta\bar{\kappa}}{\myvar{\tau}} 
  + \myconst{a}{\alpha,\beta}
    \myvar{\tau} 
    \myShiftedB{\alpha\beta\bar{\kappa}}{\myvar{\tau}}, 
\end{equation}
where $\bar\kappa$ indicates the inverse shift, 
\begin{equation}
  \myShift{\bar{\kappa}} 
  = 
  \myShift{\kappa}^{-1}, 
  \quad 
  \myShift{\alpha\bar{\kappa}} 
  = 
  \myShift{\alpha}\myShift{\kappa}^{-1}
  \quad
  \mbox{etc}, 
\end{equation}
when rewritten in terms of $\myvar{\mytauV_{\kappa}}$ and 
$\myvar{\mytauU_{\kappa}}$ given by \eref{def:sr-3}, 
are exactly \eref{eq:kappa-s} and \eref{eq:kappa-r}. 

To prove \eref{eq:kappa-s} and \eref{eq:kappa-r} for the functions defined 
in \eref{def:sr-2} 
as well as \eref{eq:kappa-t} 
we need some identities following from \eref{syst:hal} which we derive now. 

Consider the quantities 
$\mathfrak{s}_{\alpha,\beta}$, 
$\mathfrak{r}_{\alpha,\beta}$ and 
$\mathfrak{t}_{\alpha,\beta}$ which generate equations \eref{syst:hal}, 
\mynumparts{}
\begin{eqnarray}
    \mathfrak{s}_{\alpha,\beta} 
  & = & 
    \myShiftedB{\alpha}{\myvar{\sigma}} 
    \myShiftedB{\beta}{\myvar{\tau}} 
  - \myShiftedB{\alpha}{\myvar{\tau}} 
    \myShiftedB{\beta}{\myvar{\sigma}} 
  - \myconst{a}{\alpha,\beta}
    \myvar{\tau} 
    \myShiftedB{\alpha\beta}{\myvar{\sigma}} 
\\
    \mathfrak{r}_{\alpha,\beta} 
  & = & 
    \myShiftedB{\alpha}{\myvar{\tau}} 
    \myShiftedB{\beta}{\myvar{\rho}} 
  - \myShiftedB{\alpha}{\myvar{\rho}} 
    \myShiftedB{\beta}{\myvar{\tau}} 
  - \myconst{a}{\alpha,\beta}
    \myvar{\rho} 
    \myShiftedB{\alpha\beta}{\myvar{\tau}} 
\\
    \mathfrak{t}_{\alpha,\beta} 
  & = & 
    \myvar{\rho} 
    \myShiftedB{\alpha\beta}{\myvar{\sigma}} 
  + \myvar{\tau} 
    \myShiftedB{\alpha\beta}{\myvar{\tau}} 
  - \myconst{b}{\alpha\beta}
    \myShiftedB{\alpha}{\myvar{\tau}} 
    \myShiftedB{\beta}{\myvar{\tau}} 
\end{eqnarray}
\myendnumparts
By simple algebra one can obtain that 
\mynumparts{}
\begin{eqnarray}
  - \myvar{\rho} 
    \myShiftedB{\gamma}{\mathfrak{s}_{\alpha,\beta}} 
  + \myShiftedB{\beta\gamma}{\myvar{\tau}} 
    \mathfrak{t}_{\alpha,\gamma} 
  - \myShiftedB{\alpha\gamma}{\myvar{\tau}} 
    \mathfrak{t}_{\beta,\gamma} 
  & = & 
    \myShiftedB{\gamma}{\myvar{\tau}} 
    \mathfrak{T}_{\alpha,\beta,\gamma} 
\\
  - \myvar{\tau} 
    \myShiftedB{\gamma}{\mathfrak{s}_{\alpha,\beta}} 
  + \myShiftedB{\alpha\gamma}{\myvar{\sigma}} 
    \mathfrak{t}_{\beta,\gamma} 
  - \myShiftedB{\beta\gamma}{\myvar{\sigma}} 
    \mathfrak{t}_{\alpha,\gamma} 
  & = & 
    \myShiftedB{\gamma}{\myvar{\tau}} 
    \mathfrak{S}_{\alpha,\beta,\gamma} 
\\
  - \myShiftedB{\alpha\beta\gamma}{\myvar{\tau}} 
    \mathfrak{r}_{\alpha,\beta} 
  + \myShiftedB{\beta}{\myvar{\rho}} 
    \myShiftedB{\alpha}{\mathfrak{t}_{\beta,\gamma}} 
  - \myShiftedB{\alpha}{\myvar{\rho}} 
    \myShiftedB{\beta}{\mathfrak{t}_{\alpha,\gamma}} 
  & = & 
    \myShiftedB{\alpha\beta}{\myvar{\tau}} 
    \mathfrak{R}_{\alpha,\beta,\gamma} 
\end{eqnarray}
\myendnumparts 
where 
\mynumparts{}
\begin{eqnarray}
    \mathfrak{T}_{\alpha,\beta,\gamma} 
  & = & 
    \myconst{a}{\alpha,\beta}
    \myvar{\rho} 
    \myShiftedB{\alpha\beta\gamma}{\myvar{\sigma}} 
  - \myconst{b}{\alpha\gamma}
    \myShiftedB{\alpha}{\myvar{\tau}} 
    \myShiftedB{\beta\gamma}{\myvar{\tau}} 
  + \myconst{b}{\beta\gamma}
    \myShiftedB{\beta}{\myvar{\tau}} 
    \myShiftedB{\alpha\gamma}{\myvar{\tau}} 
\label{def:app-T} 
\\
    \mathfrak{S}_{\alpha,\beta,\gamma} 
  & = & 
    \myconst{a}{\alpha,\beta}
    \myvar{\tau} 
    \myShiftedB{\alpha\beta\gamma}{\myvar{\sigma}} 
  - \myconst{b}{\beta\gamma}
    \myShiftedB{\beta}{\myvar{\tau}} 
    \myShiftedB{\alpha\gamma}{\myvar{\sigma}} 
  + \myconst{b}{\alpha\gamma}
    \myShiftedB{\alpha}{\myvar{\tau}} 
    \myShiftedB{\beta\gamma}{\myvar{\sigma}} 
\label{def:app-S} 
\\
    \mathfrak{R}_{\alpha,\beta,\gamma} 
  & = & 
    \myconst{a}{\alpha,\beta}
    \myvar{\rho} 
    \myShiftedB{\alpha\beta\gamma}{\myvar{\tau}} 
  - \myconst{b}{\beta\gamma}
    \myShiftedB{\beta}{\myvar{\rho}} 
    \myShiftedB{\alpha\gamma}{\myvar{\tau}} 
  + \myconst{b}{\alpha\gamma}
    \myShiftedB{\alpha}{\myvar{\rho}} 
    \myShiftedB{\beta\gamma}{\myvar{\tau}} 
\label{def:app-R} 
\end{eqnarray}
\myendnumparts 
and then that 
\begin{equation} 
\fl\qquad
    \myconst{a}{\gamma,\delta}
    \myvar{\rho} 
    \myShiftedB{\delta}{\mathfrak{T}_{\alpha,\beta,\gamma}} 
  + \myconst{b}{\alpha\gamma}
    \myShiftedB{\alpha\delta}{\myvar{\tau}} 
    \mathfrak{R}_{\gamma,\delta,\beta} 
  - \myconst{b}{\beta\gamma}
    \myShiftedB{\beta\delta}{\myvar{\tau}} 
    \mathfrak{R}_{\gamma,\delta,\alpha} 
  = 
    \myShiftedB{\delta}{\myvar{\rho}} 
    \mathfrak{X}_{\alpha,\beta,\gamma,\delta} 
\end{equation}
where 
\begin{equation}
\fl\qquad
    \mathfrak{X}_{\alpha,\beta,\gamma,\delta} 
  = 
    \myconst{a}{\alpha,\beta}
    \myconst{a}{\gamma,\delta}
    \myvar{\rho} 
    \myShiftedB{\alpha\beta\gamma\delta}{\myvar{\sigma}} 
  - \myconst{b}{\alpha\gamma}
    \myconst{b}{\beta\delta}
    \myShiftedB{\alpha\delta}{\myvar{\tau}} 
    \myShiftedB{\beta\gamma}{\myvar{\tau}} 
  + \myconst{b}{\alpha\delta}
    \myconst{b}{\beta\gamma}
    \myShiftedB{\alpha\gamma}{\myvar{\tau}} 
    \myShiftedB{\beta\delta}{\myvar{\tau}}. 
\label{def:app-X} 
\end{equation}
These identities immediately imply that 
\begin{equation}
  \left\{ 
  \begin{array}{l}
	\mathfrak{s}_{\alpha,\beta} = 0 \\
  \mathfrak{r}_{\alpha,\beta} = 0 \\
  \mathfrak{t}_{\alpha,\beta} = 0 
  \end{array}
  \right.
  \quad\Rightarrow\quad 
  \left\{ 
  \begin{array}{l}
  \mathfrak{T}_{\alpha,\beta,\gamma} = 0 \\
  \mathfrak{S}_{\alpha,\beta,\gamma} = 0 \\
  \mathfrak{R}_{\alpha,\beta,\gamma} = 0 
  \end{array}
  \right.
  \quad\Rightarrow\quad 
  \mathfrak{X}_{\alpha,\beta,\gamma,\delta} = 0 
\end{equation}

Replacing $\gamma\to\kappa$ in \eref{def:app-S} and \eref{def:app-R} 
and applying 
$\myShift{\kappa}^{-1}$ to the latter 
one arrives at the following result: solutions for 
system \eref{syst:hal} 
solve 
\mynumparts{}
\begin{eqnarray}
    \myconst{a}{\alpha,\beta}
    \myvar{\tau} 
    \myShiftedB{\alpha\beta\kappa}{\myvar{\sigma}} 
  & = & 
    \myconst{b}{\beta\kappa}
    \myShiftedB{\beta}{\myvar{\tau}} 
    \myShiftedB{\alpha\kappa}{\myvar{\sigma}} 
  - \myconst{b}{\alpha\kappa}
    \myShiftedB{\alpha}{\myvar{\tau}} 
    \myShiftedB{\beta\kappa}{\myvar{\sigma}}, 
\\
    \myconst{a}{\alpha,\beta}
    \myShiftedB{\bar{\kappa}}{\myvar{\rho}} 
    \myShiftedB{\alpha\beta}{\myvar{\tau}} 
  & = & 
    \myconst{b}{\beta\kappa}
    \myShiftedB{\alpha}{\myvar{\tau}} 
    \myShiftedB{\beta\bar{\kappa}}{\myvar{\rho}} 
  - \myconst{b}{\alpha\kappa}
    \myShiftedB{\beta}{\myvar{\tau}} 
    \myShiftedB{\alpha\bar{\kappa}}{\myvar{\rho}} 
\end{eqnarray}
\myendnumparts 
which, rewritten in terms of 
$\myvar{\mytauQ_{\kappa}}$ and 
$\myvar{\mytauR_{\kappa}}$ given by \eref{def:sr-2}, 
are equations \eref{eq:kappa-s} and \eref{eq:kappa-r}. 

In a similar way, 
equation 
$\mathfrak{X}_{\alpha,\kappa,\beta,\kappa} = 0$ 
after the application of $\myShift{\kappa}^{-1}$ yields
\begin{equation}
\fl\qquad 
    \myconst{a}{\alpha,\kappa}
    \myconst{a}{\beta,\kappa}
    \myShiftedB{\bar{\kappa}}{\myvar{\rho}} 
    \myShiftedB{\alpha\beta\kappa}{\myvar{\sigma}} 
  + \myconst{b}{\alpha\kappa}
    \myconst{b}{\beta\kappa}
    \myShiftedB{\kappa}{\myvar{\tau}} 
    \myShiftedB{\alpha\beta\bar{\kappa}}{\myvar{\tau}} 
  = 
    \myconst{b}{\kappa\kappa}
    \myconst{b}{\alpha\beta}
    \myShiftedB{\alpha}{\myvar{\tau}} 
    \myShiftedB{\beta}{\myvar{\tau}}. 
\end{equation}
Using \eref{def:sr-2} and \eref{def:sr-3}, one arrives at 
\eref{eq:kappa-t}.

This concludes the proof of Proposition \ref{prop:kappa}.


\end{document}